\documentclass[prl,twocolumn,showpacs,superscriptaddress,preprintnumbers
,amssymb,longbibliography]{revtex4-2}
\usepackage{graphicx}
\usepackage{float}
\usepackage{color}
\usepackage{dcolumn}
\usepackage{bm}
\usepackage{physics}
\usepackage{hyperref}

\newcommand{\beq}{\begin{equation}}
\newcommand{\eeq}{\end{equation}}
\newcommand{\beqn}{\begin{eqnarray}}
\newcommand{\eeqn}{\end{eqnarray}}

\renewcommand{\vec}[1]{\boldsymbol{#1}}

\newcommand{\vect}[1]{{\bm{#1}}}

\begin{document}

\title{Phase Transition of Topological Index driven by Dephasing}

\author{Thomas G. Kiely}

\affiliation{Kavli Institute for Theoretical Physics, University of California, Santa Barbara, CA 93106, USA}

\author{Cenke Xu}

\affiliation{Department of Physics, University of California,
Santa Barbara, CA 93106, USA}

\date{\today}

\begin{abstract}

We study topological insulators under dephasing noise. With examples of both a $2d$ Chern insulator and a $3d$ topological insulator protected by time-reversal symmetry, we demonstrate that there is a phase transition at finite dephasing strength between phases with nontrivial and trivial topological indices. Here the topological index is defined through the correlation matrix. The transition can be diagnosed through the spectrum of the whole correlation matrix or of a local subsystem. Interestingly, even if the topological insulator is very close to the topological-trivial critical point in its Hamiltonian, it still takes finite strength of dephasing to change the topological index, suggesting the robustness of topological insulators under dephasing. {
We further consider Chern insulators in the presence of real-space disorder, which exhibit a ground-state transition between topological and Anderson insulating phases. We find that even strongly-disordered Chern insulators, close to the critical disorder strength, exhibit robustness with respect to dephasing.}


\end{abstract}

\maketitle

{\bf --- Introduction}

The study of open quantum systems far from equilibrium is an emerging field with strong connections to modern-day theories and to experiments on quantum materials
and quantum simulators. Fascinating physics in this direction have been uncovered, including Floquet states of matter~\cite{tcrystal1,tcrystal2,tcrystal3,tcrystal4,tcrystal5,tcrystal6,floquetcode1,floquetcode2,floquetcode3,floquetcode4,floquetcode5,fspt1,fspt2,fspt3,fspt4,fspt5,fspt6,fspt7,fspt8}, measurement-induced phase transitions~\cite{MIPT1,MIPT2,MIPTtheory1,MIPTtheory2}, measurement altered criticality~\cite{maojian,zhouhsieh,aliceaising}, strong-to-weak spontaneous symmetry breaking~\cite{wfdecohere,biswssb,biswssb2,yizhiswssb}, classification of Gaussian mixed-state topological insulators~\cite{cooper,openfermion}, as well as mixed state symmetry protected topological phases~\cite{de_Groot_2022,mawang,sptdecohere,qibi,huang2024mixedstatetopologicalorder,huang2025topologicalresponseopenquantum}. When discussing open quantum systems, the foremost question to address is whether a quantum state can maintain its pure state information when exposed to an environment.
The Lieb-Robinson bound dictates that long distance correlations
between local operators, ${\rm Tr}(\rho A(x) B(y))$, cannot change qualitatively under finite-depth decoherence or finite-time Lindbladian evolution.
Topological transitions, however, do not require a fundamental change of long distance local correlations. Hence it is in principle possible to find transitions of topological nature under decoherence. The most well-known example along this line is toric code topological order under dephasing: there is a critical strength of dephasing beyond which the quantum information stored in the toric code becomes irretrievable under decoding. Beyond this threshold, the quantum nature of the toric code is lost~\cite{kitaevpreskill,wfdecohere,altman2,fan2023}.



Here we discuss topological insulators (TI) in an open environment. In previous works the authors have investigated Chern insulators under partial or full dephasing with different methods, including field theory, R\'{e}nyi correlators, relative R\'{e}nyi entropy, and ``strange correlators"~\cite{cherndephase2,choiSL}. 
These methods provide insight into the structure of the dephased density matrices, without exploring the possible topological transitions caused by dephasing.
In this work we will discuss the possibility of a topological transition driven by dephasing. Moreover, the transition we find can be identified using the experimental observables present in ultracold atom quantum simulators.

We first need to define what we mean by topological transition at finite dephasing. For a pure quantum system, there are various ways of defining the topological indices. The Chern number of a noninteracting Chern insulator, for example, can be computed using the Bloch state wave functions and Berry phase, free fermion Green's function, response to external electromagnetic field as well as twisted boundary conditions, etc. When the system is interacting, some of these methods fail -- for example, the single particle Berry phase is no longer well-defined. But if one can compute the response to external EM field (the Hall conductance), the many-body Chern number can still be inferred. 

Dephasing noise can be formulated as an interaction at the temporal interface of a Euclidean space-time in path integral~\cite{altman2}. This interaction invalidates Wick's theorem, which means that the dephased system generically constitutes an ``interacting" problem.
From another perspective, dephasing noise can be interpreted as
the coupling
of a system to ancilla qubits through Kraus operators.
By expressing this coupling as dephasing noise, however, we 
have ``integrated out"
the real-time dynamics of the 
ancilla qubits. This inhibits the computation of the Hall conductance and effective Chern-Simons theory,
as these procedures involve calculations at finite frequency. 

In this work we define topological indices through correlation matrix, $C_{ij}^{\alpha\beta}={\rm Tr}(\rho c^\dagger_{i,\alpha}c_{j,\beta})$ where $i,j$ are index positions and $\alpha,\beta$ denote internal (orbital and/or spin) degrees of freedom. The same diagnosis of ``topology" has been used in previous 
works~\cite{disTI1,disTI2,disTI3,disTI5,disTI6}. For a non-interacting topological insulator, $C_{ij}^{\alpha\beta}$ is a projection operator onto the filled band. Alternatively, it can be thought of as a topologically-equivalent Hamiltonian with identical eigenstates but perfectly flat bands~\cite{turner2010}. This ``spectral-flattening" trick has been used to compute eigenvalues of reduced density matrices~\cite{Peschel_2003,turner2010,Fidkowski2010} and to define topological states via winding numbers~\cite{hannukainen2024,lippo2025} and spectral flow~\cite{turner2010,hughes2011,alexandradinata2011,turner2012}. It is closely related to the Green's function definition of topological invariants~\cite{wang2012,wang2012b} for interacting TIs. Here we show that all these diagnostics provide a consistent picture of robust topological order in the presence of dephasing. Furthermore, the real-space correlation matrix is in principle observable in experiments using ultracold atoms or superconducting qubits~\cite{knap2013,pena2018,impertro2024}. The occupation number of the momentum-space eigenstates of the correlation matrix can also be measured through the time-of-flight technique~\cite{bloch2008,alba2011,rubiogarcia2020}.


The evolution we consider here is governed by the Lindblad master equation,
\begin{equation}\label{eq:lindblad}
{
\partial_g\rho=\sum_{l}\left(L_{l}\rho L^\dagger_{l}-\frac{1}{2}\left\{\rho,L^\dagger_{l}L_{l} \right\}\right).}
\end{equation}
{
As we are interested in the topological features of the dephased wavefunction, we do not consider any Hamiltonian evolution.}
Density dephasing is defined by choosing the Lindblad jump operators $L_i$ to be local density operators, $n_{l}$, on site $l$. There is a physically-relevant choice about how to include internal degrees of freedom (e.g. spin and orbital indices, denoted by $\alpha$) in Eq.~(\ref{eq:lindblad}): one can define $L_l=\sum_\alpha n_{l,\alpha}$ or one can define distinct jump operators acting on each internal state, $L_{l,\alpha}=n_{l,\alpha}$. The choice one makes in this regard derives from how dephasing noise (e.g. ancilla degrees of freedom) couples to the states on particular sites. Here we choose the former, which assumes that the ancilla coupling is independent of the internal state{
, as this is more relevant for applications in analog quantum simulators.}


The dephasing strength is parameterized by $g$: $\rho(g=0)$ is a pure state TI, while $\rho(g\to\infty)$ is a fully-dephased state characterized by a diagonal correlation matrix.
Remarkably, although $\rho(g>0)$ is no longer Gaussian, the dephased correlation matrix obeys a simple equation of motion~\cite{dolgirev2020,turkeshi2021}:
\begin{equation}\label{eq:dC}{
    \partial_gC_{ij}^{\alpha\beta}=-\left(1-\delta_{ij}\right)C_{ij}^{\alpha\beta}
    }
\end{equation}

The correlation matrix of the dephased state is no longer a projection operator -- its eigenvalues will take values between $0$ and $1$. With that said, we may extend the definition of the flat-band Hamiltonian $H_C=\mathbb{I}/2-C$~\cite{turner2010} to finite dephasing and study the properties of its eigenstates as a function of $g$. We will use this as a tool to characterize the topology of the dephased state. As $g\to 0^+$, the ground state of $H_C$ should have the same topological order as that of $H$. As $g\to\infty$, the correlation matrix $C$ will contain only diagonal elements and the ground-state of $H_C$ should be trivial.


{\bf --- Dephased Chern Insulator}

We consider a 2D Chern insulator on a square lattice defined by the following Bloch Hamiltonian~\cite{QWZ}: \beqn\label{eq:2dti} H(k_x, k_y) &=& \sin(k_x) \sigma^x + \sin(k_y) \sigma^y \cr\cr &+& (\gamma - \cos(k_x) - \cos(k_y)) \sigma^z. \eeqn
This model hosts gapped ground states with Chern number {
$C=-1$ for $-2<\gamma<0$ and $C=+1$ for $0<\gamma<2$.} The ground state is trivial for $|\gamma|>2$. 

\begin{figure}
    \centering
    \includegraphics[width=\linewidth]{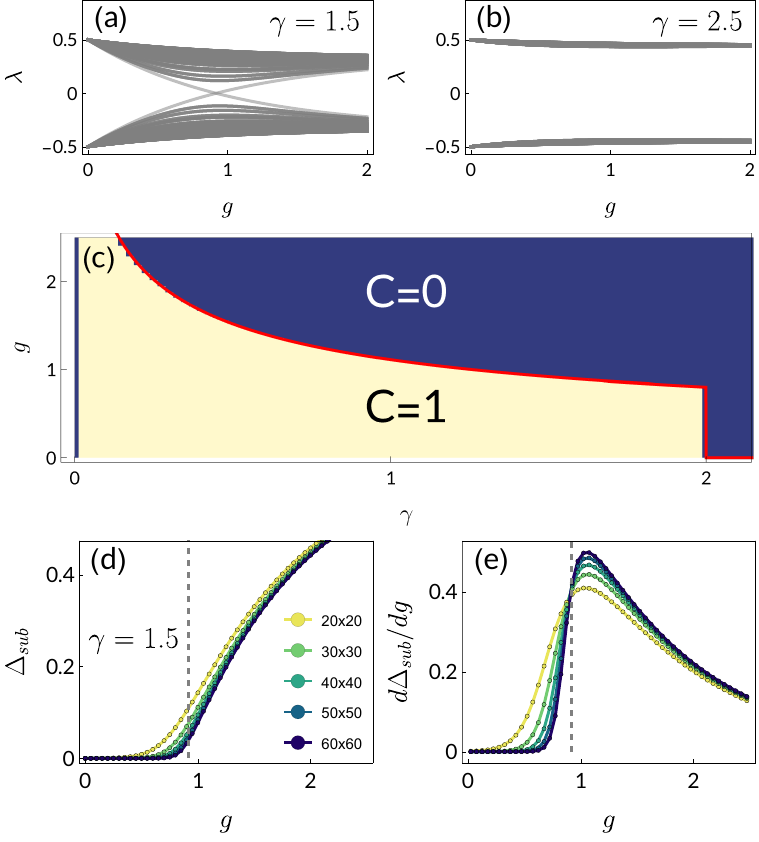}
    \caption{Dephasing-driven topological transition in a Chern insulator. (a) Evolution of the spectrum of $H_C=\mathbb{I}/2-C$ as a function of dephasing strength $g$ in the topological phase ($\gamma=1.5$). A gap closes in the spectrum at a critical dephasing strength $g_c$.
    (b) In the trivial phase ($\gamma=2.5$), the upper and lower bands of the correlation matrix remain separated throughout the evolution. (c) Phase diagram of the dephased Chern insulator on a $12\times12$ torus. Beige region denotes the topological phase with Chern number $C=1$. Navy region denotes the trivial phase ($C=0$).
    Red line is numerical solution for $g_c$ in the thermodynamic limit (see main text).
    The critical dephasing strength remains finite as $\gamma\to 2$ and diverges as $\gamma\to 0$. (d) Correlation sub-matrix gap $\Delta_{sub}$ on tori of various sizes. The gap is zero in the topological phase and finite in the trivial phase (see main text). Gray dashed line shows the critical dephasing strength in the thermodynamic limit{
    , $g_c\approx0.918$.} (e) The derivative of the sub-matrix gap with respect to $g$ for the same system sizes. The curves intersect at the critical dephasing strength.
    }
    \label{fig:2dti_chern}
\end{figure}

In Fig.~\ref{fig:2dti_chern}a we show the evolution of the correlation matrix spectrum as a function of dephasing strength $g$ in the topological phase ($\gamma=1.5$). Results are shown for a $20\times20$ torus. At a critical dephasing strength $g_c$, there is a level crossing at momentum $\vec k=0$. As the system size is enlarged, we find an emerging continuum of low-energy modes in the vicinity of $g_c$. In momentum space, these modes form a Dirac cone centered at $\vec k=0$ at the critical dephasing $g=g_c$.
In Fig.~\ref{fig:2dti_chern}b we compare this 
to the evolution of the trivial phase, where the upper and lower bands remain well separated for all $g$.

As the Lindblad evolution is spatially uniform, momentum remains a good quantum number. Thus we can define Bloch eigenstates of $H_C$ and integrate their Berry curvature to determine the Chern number of the low-energy band. In Fig.~\ref{fig:2dti_chern}c we show the Chern number as a function of $\gamma$ and $g$ on a $12\times12$ torus. Indeed we find that the gap closing in Fig.~\ref{fig:2dti_chern}a is associated with a sharp change in the Chern number, and thus can be considered a topological phase transition. Notably, we find that the Chern insulator is surprisingly robust to dephasing noise. Although energetic gaps close at the points $\gamma=2$ and $\gamma=0$, the critical dephasing strength $g_c$ does not vanish at these points. As $\gamma\to 0$ we {
in fact} find that $g_c$ diverges logarithmically {
due to a band-inversion -- the correlation matrix gap closes at the Brillouin zone corners for $\gamma<0$}.

The red line in Fig.~\ref{fig:2dti_chern}c shows a numerical calculation of the critical dephasing strength $g_c$ as a function of $\gamma$ in the thermodynamic limit. This is determined by rewriting Eq.~(\ref{eq:dC}) in momentum space and taking the continuum limit:
\begin{equation}\label{eq:dCk}
    \partial_gC^{\alpha\beta}(\vec k)=-C^{\alpha\beta}(\vec k)+\int\frac{d^2 k^\prime}{(2\pi)^2}C^{\alpha\beta}(\vec k^\prime).
\end{equation}
By integrating the equation of motion over $\vec k$, one finds that the last term is a constant of motion.
Given the value of this term and the initial $C^{\alpha\beta}(\vec k)$, we can solve the differential equation explicitly. 
At $\vec k=0$, where the Dirac cone closes, the correlation matrix is diagonal: $C(\vec k,g)=d_0(g)\mathbb{I}+d_1(g)\sigma^z$. The topological transition is identified by $d_1(g)$ changing sign. Solving for the point where $d_1(g)$ vanishes yields the red line in Fig.~\ref{fig:2dti_chern}c.

At the critical dephasing, $g_c$, we emphasize that the correlation matrix in real space is still short-ranged. In momentum space, the correlation matrix has ``zeros" at $\vect{k} = 0$: $C(\vec k=0, g_c) - d_0(g_c) \mathbb{I}=0$. This is similar to other transitions of topological indices observed before in interacting systems with topological bands~\cite{zero1,zero2,SMG1}.  

Another natural quantity to use in characterizing the topological transition is the spectrum of $C^{\alpha\beta}_{ij\in A}$, a sub-matrix of the real-space correlation matrix in which sites $i$ and $j$ are confined to a particular region $A$. If $A$ is a cylinder with open boundaries, even if the full system has no boundaries, the spectrum of the sub-matrix will exhibit spectral flow (``charge pumping") when flux is threaded through the cylinder. For Gaussian states, this spectral flow can be explicitly understood in terms of the Li-Haldane conjecture~\cite{li2008} by expressing the reduced density matrix in region $A$ in terms of $C^{\alpha\beta}_{ij\in A}$~\cite{Peschel_2003,Fidkowski2010}. In Fig.~\ref{fig:2dti_chern}d we plot the spectral gap $\Delta_{sub}$ of the sub-matrix, defined as the difference between the largest eigenvalue less than 1/2 and the smallest eigenvalue greater than 1/2,
as a function of $g$. Different curves refer to different system sizes; in all cases, region $A$ and its complement constitute a bipartition of the torus. The spectral gap is zero in the topological phase, which is directly related to the gapless spectrum of the TI on an open cylinder~\cite{turner2010}. For dephasings $g\geq g_c$, a gap opens in the submatrix spectrum.
While convergence with respect to system size is slow, we show in Fig.~\ref{fig:2dti_chern}e that intersection of $d\Delta_{sub}/dg$ for different system sizes can be used to extract the critical dephasing strength.



{\bf --- Dephased $3d$ TI}

One can ask whether the above topological transition was specific to a single filled Chern band. Here we apply the same analysis to a three-dimensional topological insulator protected by time-reversal symmetry~\cite{fu2007}.
The Bloch Hamiltonian for this model takes the form
\begin{eqnarray}\label{eq:3dti_h}
    H(\vec k)&=& \sum_{i=1}^5h_i(\vec k)\Gamma_i
\end{eqnarray}
where the $\Gamma_i$ are Dirac matrices, 
\begin{equation}
    \Gamma_{(1,2,3,4,5)}=\left(\sigma^x\otimes\mathbb{I},\sigma^y\otimes\mathbb{I},\sigma^z\otimes s^x,\sigma^z\otimes s^y,\sigma^z\otimes s^z\right).
\end{equation}
The coefficients are
\begin{eqnarray*}
    h_1(\vec k)&=&t+\delta t+t(\cos(u_1)+\cos(u_2)+\cos(u_3)) \\
    h_2(\vec k)&=&t(\sin(u_1)+\sin(u_2)+\sin(u_3)) \\
    h_3(\vec k)&=&\lambda(\sin(u_2)-\sin(u_3)-\sin(u_2-u_1)+\sin(u_3-u_1)) \\
    h_4(\vec k)&=&\lambda(\sin(u_3)-\sin(u_1)-\sin(u_3-u_2)+\sin(u_1-u_2)) \\
    h_5(\vec k)&=&\lambda(\sin(u_1)-\sin(u_2)-\sin(u_1-u_3)+\sin(u_2-u_3))
\end{eqnarray*}
where $u_i=\vec k\cdot\vec a_i$ with $\vec a_1=(0,1,1)/2$, $\vec a_2=(1,0,1)/2$ and $\vec a_3=(1,1,0)/2$. This model corresponds to a tight-binding model on a diamond lattice where the tunneling matrix element is distorted along the $[111]$ direction. The tunneling amplitude along this direction is $t+\delta t$. For $\lambda\neq 0$, this model hosts both weak and strong TI phases depending on the value of $\delta t$~\cite{fu2007}.
In the following, we set $t=1$ and $\lambda=1/8$.

\begin{figure}
    \centering
    \includegraphics[width=\linewidth]{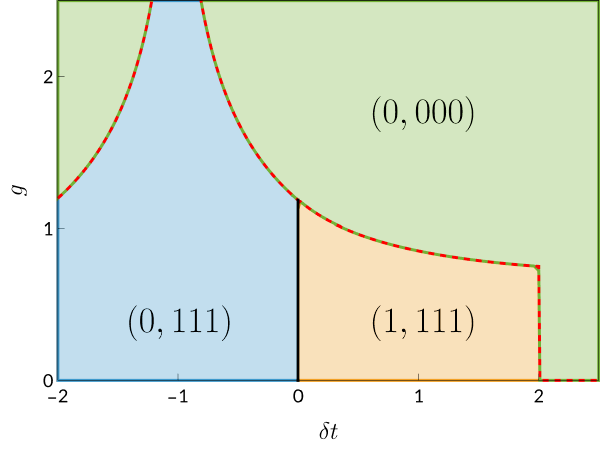}
    \caption{Dephasing-driven topological transition in a 3D TI.
    The model is given by Eq.~(\ref{eq:3dti_h}) with $t=1$ and $\lambda=1/8$. The orange region is the strong TI phase with topological invariants $(\nu_0,\nu_1\nu_2\nu_3)=(1,111)$. The blue region is the weak TI phase with invariants $(0,111)$. {
    They are separated by a gapless line at $\delta t=0$ (black).} Both phases transition to a trivial phase (green) above some critical $g_c$, {
    shown with a red dashed line}. 
    Note that $g_c$ is finite as $\delta t\to 2^-$ and diverges as $\delta t\to-1$.}
    \label{fig:3dti}
\end{figure}

We find that both the strong and weak TI phases are robust to dephasing noise. In the strong TI phase, the dephased correlation matrix exhibits a gap closing at the $T$ point, $\vec k_T=(\pi/2,\pi/2,\pi/2)$. 
As this is a time-reversal-invariant momentum point (i.e. $\vec k_T=-\vec k_T$ modulo a reciprocal lattice vector), the correlation matrix at $\vec k_T$ takes the form $C(\vec k_T,g)=d_0(g)\mathbb{I}+d_1(g)\Gamma_1$~\cite{fu2007}. As with the Chern insulator, the topological phase transition corresponds to $d_1(g)$ changing sign. This leads to a change in the strong topological invariant as defined in Ref.~\cite{fu2007}. 

{
In the weak TI phase with $-1<\delta t<0$, the critical point is characterized by 4 Dirac cones (at the $T$ point and $X$ points in the Brillouin zone~\cite{fu2007}). Another weak TI phase is present for $-2<\delta t<-1$ in which the tunneling along $[111]$ changes sign and the 4 Dirac cones shift to the $\Gamma$ and $L$ points, respectively.
}
The presence of an even number of Dirac cones on boundary surfaces is precisely what prevents the weak TI phase from being robust to disorder.

With these conditions, we numerically integrate the last term in Eq.~(\ref{eq:dCk}) and solve for the critical dephasing strength $g_c$ as a function of $\delta t$. Our results are shown in Fig.~\ref{fig:3dti}. As with the Chern insulator, we find that the 3D topological insulator is robust to weak dephasing, even arbitrarily close to the critical points at $\delta t=0,2$. The critical dephasing strength $g_c$ diverges logarithmically at the point $\delta t=-1$, where the tunneling amplitude along $[111]$ vanishes.

{\bf --- Disordered Chern insulator}

We have shown that clean TIs are surprisingly robust against dephasing, even arbitrarily close to the trivial-topological phase transition. Here we show that this robustness is not particular to this kind of topological transition. We consider the model of a Chern insulator, Eq.~(\ref{eq:2dti}), in the presence of random on-site disorder. For sufficiently strong disorder, this model hosts a
Chern-Anderson transition between a topological Chern insulator and a trivial Anderson insulator~\cite{MORENOGONZALEZ2023169258}. 

As disorder breaks translational symmetry, our analytical arguments can no longer be used to establish the phase boundaries. In order to establish the topology of the disordered state, we treat a finite-size periodic system as if it were a unit-cell embedded in an infinitely large array. In this interpretation, one can probe different points in the Brillouin zone by twisting the boundary condition on the torus, allowing us to integrate the Berry curvature and define a Chern number~\cite{fukui2005,prodan2010,Zhang_2013}. In the absence of disorder, this is identical to our momentum-space calculation. The topology at fixed disorder strength is defined by the modal outcome: do more states have $C=0$ or $C\neq 0$? For our finite system, if for a typical disorder realization the system has a nonzero $C$, then in the thermodynamics limit the $C \neq 0$ domains are expected to percolate and drive the entire system into a Chern insulator. Thus, this transition is expected to become sharp in the thermodynamic limit~\cite{MORENOGONZALEZ2023169258}. 

\begin{figure}
    \centering
    \includegraphics[width=\linewidth]{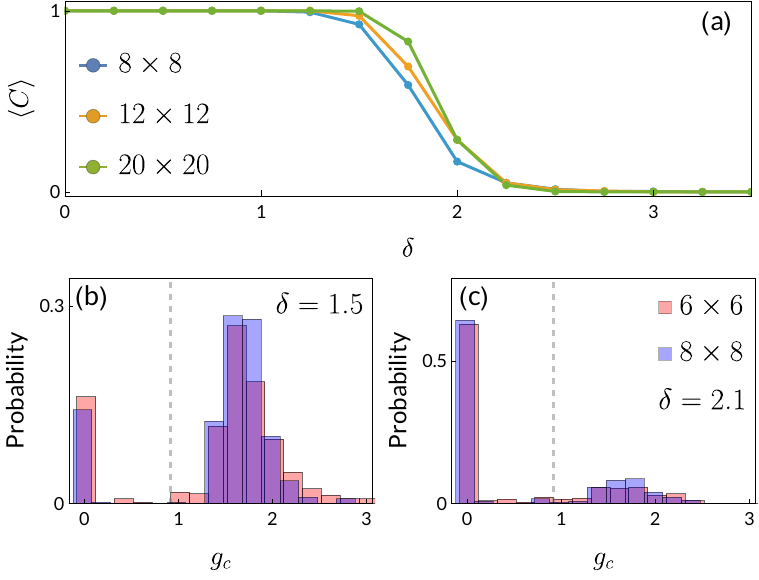}
    \caption{Dephased and disordered Chern insulator. (a) Average Chern number as a function of disorder width $\delta$ for tori of various sizes. Average is taken over 400 disorder realizations. The Chern-Anderson transition is smooth for finite system sizes but sharpens as the system size grows. (b-c) Distribution of critical dephasing strengths $g_c$ for different disorder configurations at fixed disorder widths $\delta$ and $\gamma=1.5$. Orange data is on a $6\times6$ torus and blue data is $8\times8$. Gray dashed line is $g_c$ for $\gamma=1.5$ and $\delta=0$. (a) For $\delta=1.5<\delta_c$, we see the modal $g_c>0$. (b) For $\delta=2.1>\delta_c$, the modal $g_c=0$. In both cases, the distribution is bimodal. This feature tells us that even highly disordered Chern insulators are quite robust to dephasing. }
    \label{fig:2dti_disorder}
\end{figure}

We study Eq.~(\ref{eq:2dti}) with Gaussian disorder parameterized by its width $\delta$. In Fig.~\ref{fig:2dti_disorder}a we plot the average Chern number as a function of $\delta$ over $400$ disorder realizations at $\gamma=1.5$. We find that the average Chern number smoothly decreases for finite-sized systems, but the transition becomes progressively sharper as system size is increased. The phase boundary $\delta_c\approx1.9$ is in rough agreement with prior results obtained by different means~\cite{MORENOGONZALEZ2023169258}. In Figs.~\ref{fig:2dti_disorder}b and \ref{fig:2dti_disorder}c we plot histograms of the critical dephasing strengths $g_c$ for the ensemble of states with $\delta=1.5<\delta_c$ and $\delta=2.1>\delta_c$. Crucially, we find that although the modal outcome shifts from $g_c>0$ to $g_c=0$ as $\delta$ is increased, the modal $g_c$ does not continuously approach zero as $\delta\to\delta_c$. This is evidenced by the bimodal distribution,
which sharpens as system size is increased.
Thus, we can conclude that even extremely ``dirty" Chern insulators, with a high degree of disorder, are robust to weak dephasing.

{\bf --- Discussion}

In this work we have investigated the dephasing-driven transition of a topological index. The topological index is defined for the dephased mixed state through its correlation matrix, and the change in this index coincides with changes to the spectrum of the whole or subsystem correlation matrix.
We demonstrate that it always takes a finite dephasing strength to drive a transition of topological index, even when the system is tuned in proximity to a topological-trivial transition in its parent local Hamiltonian, or to a Chern-Anderson insulator transition with strong disorder. This suggests that certain topological features, such as the topological index of a system, are robust against dephasing. 

We caution the readers that different indicators of nontrivial topology do not necessarily coincide with each other. There are concrete examples of topological indices defined through single particle Green's function (hence correlation matrix) that do not generate the same transition point as other definitions~\cite{zero2,SMG1}.
The strange correlator is another numerical tool proposed to diagnose the topological nature of a pure or mixed states~\cite{strange}. In the case of dephased topological insulators, strange correlators show nontrivial behavior originating from topology all the way up to infinite dephasing strength~\cite{cherndephase2}. {
In the case of the single-particle strange correlator, this can be understood using the Lindblad-based arguments developed here: only the magnitude of the correlation function scales with $g$, so its quasi-long-ranged nature persists as $g\to\infty$.} Moreover, if one were to investigate the real boundary of the system, the edge state of a TI may have its own transition under decoherence that need not coincide with that of the bulk.

However, since various features of the correlation matrix can be measured in experiment, the transitions discussed in this
work correspond to an observable effect. For example, if the dephased topological insulator is realized in cold atom experiments {
(e.g. by averaging over a random time-varying on-site potentials)}, the spectrum of the correlation matrix can be probed through time-of-flight techniques~\cite{alba2011,rubiogarcia2020}. Alternatively, real-space correlations can be measured by engineering long-range tunnel-couplings~\cite{pena2018} or by many-body interferometry~\cite{knap2013}.
Hence, the phase diagram of topological indices under dephasing can be observed experimentally.

{
The data that support the findings of this article are openly available at~\cite{dataref}.}
Xu acknowledges support from the Simons foundation through
the Simons investigator program. TGK acknowledges support from the National Science Foundation under grant PHY-2309135 to the Kavli Institute for Theoretical Physics (KITP), and from the
Gordon and Betty Moore Foundation through Grant GBMF8690 to the University of California, Santa Barbara

\bibliography{bib}

\end{document}